\documentclass[graybox]{svmult}

\usepackage{type1cm}        
\usepackage{makeidx}         
\usepackage{graphicx}        
\usepackage{multicol}        
\usepackage[bottom]{footmisc}

\usepackage{newtxtext}       %
\usepackage{newtxmath}       

\makeindex             

\begin{document}

\title*{Globular cluster systems and Galaxy Formation}

\author{Michael A. Beasley}
\institute{Michael A. Beasley \at Instituto de Astrof\'isica de Canarias, Calle V\'ia L\'actea s/n, La Laguna, Tenerife, E-38205, Spain, \email{beasley@iac.es}}
%
%
\maketitle

\abstract{Globular clusters  are compact, gravitationally bound systems of up to $\sim1$~million stars. The GCs in the Milky Way contain some of the oldest stars known, and provide important clues to the early formation and continuing evolution of our Galaxy. More generally, GCs are associated with galaxies of all types and masses, from low-mass dwarf galaxies to the most massive early-type galaxies which lie in the centres of massive  galaxy clusters. GC systems show several properties which connect tightly with properties of their host galaxies. For example,  the total mass of GCs in a system scales linearly with the dark matter halo mass of its host galaxy. Numerical simulations are at the point of being able to resolve globular  cluster formation within a cosmological framework. Therefore, GCs  link  a range of scales, from the  physics of star formation in turbulent gas clouds, to the large-scale properties of galaxies and their dark matter. In this Chapter we review some of the basic observational approaches for GC systems, some of their key observational properties, and describe how GCs  provide important clues to the formation of their parent galaxies.
}

\section{Introduction}
\label{Introduction}

Understanding how galaxies form and evolve are major challenges in astrophysics. We presently do not have a definitive model for how nearby galaxies attain the morphologies we see, or understand fully how galaxies evolve as we look back in redshift. However, astronomers have a reasonable picture of some of the fundamental aspects of these processes. A large part of the observational and theoretical effort in understanding galaxy formation has focused on the stellar and gas content of the main components of galaxies\footnote{This is often called the "baryonic" component, since stars and gas are composed of protons and neutrons which are baryons (three quarks). The poor old electrons -- which are leptons -- are ignored in this terminology!}. In the case of early-type galaxies (e.g., elliptical galaxies; hereafter "ETGs")  this corresponds primarily to the smoothly distributed stellar body. In the case of late-type galaxies, such as the Milky Way, these components may include a stellar and gaseous disk, and perhaps a stellar bar and bulge. However, an important additional component of galaxies is their globular cluster (GC) systems. While they may only account for $<0.1$\% of the total stars in a galaxy \cite{Brodie2006}, GCs play an out-sized role in tackling several key problems in galaxy formation.

In this Chapter, some of the key properties of GC systems will be discussed, with a  particular focus on their link to the formation and evolution of their host galaxies. A number of excellent reviews on GCs and GC systems exist in the literature. For the specific case of the chemical properties of Milky Way GCs see \cite{Gratton2012} and \cite{Bastian2018}. For a review on extragalactic GC systems see \cite{Harris1991} and \cite{Brodie2006}. \cite{Forbes2018} discuss some outstanding issues in GC formation and evolution. The literature focusing specifically on galaxy formation is vast and it is not possible to go into details here. \cite{Silk2014} discuss some of the general issues in galaxy formation, while \cite{Cappellari2016} focuses on the kinematics of galaxies. See \cite{Kennicutt1998} for details on star formation in galaxies along the Hubble sequence. \cite{Renzini2006} discusses the formation and evolution of massive  ETGs. \cite{BlandHawthorn2016} go into details for the specific case of the Milky Way, while \cite{Naab2017} discuss outstanding issues in galaxy formation from a theoretical point of view. A review of dwarf galaxies can be found in \cite{McConnachie2012}, and of "ultra-faint" dwarf systems here \cite{Simon2019}. The above works are good starting points for research into GC systems and galaxy formation.

\section{What globular clusters are and are not}
\label{Whatareglobulars}

GCs are compact, centrally concentrated, gravitationally bound systems of stars.  The mean mass of a GC in the Milky Way is $\sim 2\times10^5$\,M$_\odot$, and this is typical for GC systems in general. Three parameters are  used to describe their physical sizes. The {\it core radius} ($r_{\rm c}$),  is the radius at which the central surface brightness of the cluster drops by half; the {\it half-light radius} ($r_{\rm h}$),  is  the radius that contains half the light of the cluster, and the {\it tidal radius} ($r_{\rm t}$), is the radius at which the density of the GC reaches zero \cite{King1962}. Median values for these radii in the Milky Way are $r_{\rm c}\sim1.5$\,pc, $r_{\rm h}\sim10$\,pc and $r_{\rm t}\sim50$\,pc \cite{Peterson1975,Harris1996}.

From the above numbers we see that GCs are also very dense stellar systems. GCs have average stellar densities of $\sim 0.2$\,M$_\odot$\,pc$^{-3}$ and central densities of up to $\sim 10^{4}$\,M$_\odot$\,pc$^{-3}$. These high densities and their small sizes make GCs very interesting laboratories for studies of stellar evolution and dynamics  (see \cite{Meylan1997}  for a review). However, more importantly -- from the point of view of this Chapter at least -- this also means that they are relatively luminous (mean $V$-band absolute magnitudes, $M_V\sim-7.5$), concentrated stellar sources, which makes them readily observable out to large distances. Their high densities also mean that they are quite resistant to external tidal forces; the GCs in the Milky Way have survived some 10 billion years of galaxy evolution (although many of the low- and high-mass clusters may well have been destroyed; \cite{Gnedin1997,Fall2001}. Within our ability to measure, all the stars in a GC are coeval (uniform age) and generally old. In addition, they are largely mono-metallic (same heavy-element abundance) to a high degree of uniformity. This makes GCs important laboratories for stellar evolution since the key parameter which dictates the evolutionary stage of any star in a cluster becomes its  mass. 

Extensive observational evidence suggests that galaxies are embedded in halos of "dark matter". The rotation curves of the gas in galaxies \cite{Rubin1970,Sofue2001}, the motions of the stellar component (stellar velocity dispersion) \cite{Cappellari2016}, the hot X-ray emitting gas that surrounds massive galaxies \cite{Forman1985,Humphry2006}, gravitational lensing of background sources \cite{Mandelbaum2005,Cacciato2009} and the kinematics of GC systems \cite{Strader2011,Arnold2014}, all indicate the existence of a gravitational potential in excess of that observed for the mass of  gas and stars alone. 

As far as we know, GCs {\it at the present epoch}\footnote{This qualifier is important to remember. Although GCs seem to have no dark matter now, this does not necessarily imply that they never had dark matter. Tidal processes could remove the vast majority of dark matter from a GC orbiting in a Milky Way-like potential over a Hubble time.} do not have dark matter (e.g.,  \cite{Moore1996,Meylan1997}). That is, they may contain dark stellar remnants from stellar evolution (e.g., brown dwarfs, black holes), but do not seem to possess any further invisible component comprising hypothetical exotic particles. This observation is important for a couple of reasons. Firstly, galaxies {\it do} have dark matter, so one can conclude that  GCs are not galaxies! This conclusion is more profound than it first appears; a comprehensive definition of a galaxy is not so straightforward, and definitions based on size, stellar populations or stellar density can often end up including GCs (see \cite{Willman2012})\footnote{Stellar systems of masses at, or below that of GCs are observed, and they typically have very high mass-to-light ratios ($M / L_V > 100$) implying high dark matter fractions. However, these are generally low-concentration, low-surface brightness objects which are typically larger ($r_{\rm h}> 50$\,pc) than GCs and are referred to as "ultra-faint dwarf" galaxies (UFDs; \cite{Simon2019}).}. Whether or not a system has dark matter does discriminate between galaxies and stellar clusters (globular, open or young massive clusters).

Secondly, models for GC formation may be broken down into two very broad classes; "dark matter" formation models and "baryonic" formation models. In dark matter models, GCs are required to have (at least, initially) their own dark matter halo in which to form. Initial ideas included the early collapse of gas into low-mass dark halos to form GCs \cite{Peebles1984}. Alternatively, collisions of  dark matter haloes may lead to GC formation at high redshift \cite{Madau2019}. If GCs do indeed form in their own dark matter haloes, then GCs unassociated with galaxies residing in low-density environments may exist. However, so far such "free-floating" GCs have not been observed \cite{Mackey2016}, although "unbound" GCs are seen to be present in galaxy clusters which are believed stripped from their parent galaxies (e.g., \cite{Peng2011}). 

The second class of models, baryonic models, make GC formation less special. They generally posit that GC formation is a natural extension of any clustered star formation, and that the main difference between other types of star cluster and GCs is one of mass, rather than formation mechanism \cite{Elmegreen1997,Kruijssen2015}. In this picture, GCs can also form in a dark matter haloes, but these haloes are associated with the host galaxy, rather than individual GCs. These models also make predictions; for example in many cases they predict that GCs will obey an age--metallicity relation such that  more metal-rich clusters formed in a single system may be younger than their metal-poor counterparts. Models for the formation of GC systems are further discussed in Section~\ref{Simulations}.

\section{Observational techniques}
\label{ObservationalTechniques}

The observational techniques used to study GCs can be separated into two categories. One category is in the {\it spatially resolved case}. This generally (but not exclusively) applies to "nearby" clusters  ($d < 1.0$\,Mpc), for example, GCs found in the Milky Way and other galaxies in the Local Group. In this case, individual cluster stars can be resolved and studied, and their properties measured either via photometry or spectroscopy. The second category is the {\it spatially unresolved case}. By this it is meant that individual stars are not resolved, although there may still be spatially varying information. Here, the integrated light of the cluster is studied as a sum of all the stars in the cluster. Again, the both photometry and spectroscopy can be used to obtain useful information, although the analysis techniques differ from those of  spatially resolved studies.
 
\subsection{Spatially resolved methods}
\label{SpatiallyResolved}

Spatially resolved photometry and spectroscopy are primary observational methods to obtain detailed information about the ages and chemistry of GCs. In addition, due to the proximity of Milky Way GCs, their proper motions can be measured.

\subsubsection{Colour-magnitude diagrams}
\label{CMDs}

\begin{figure}
\includegraphics[width=\textwidth]{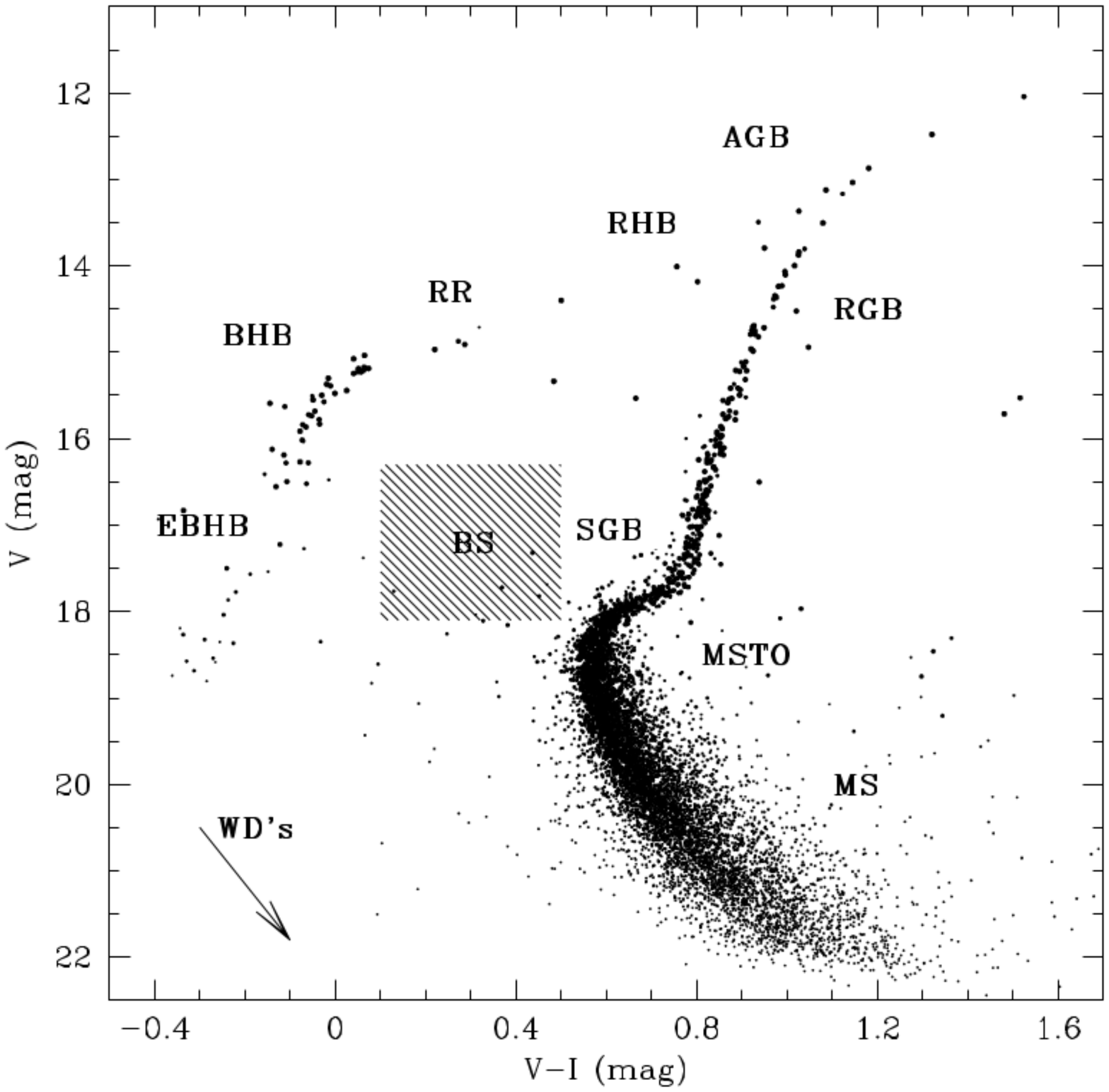}
\caption{$V,I$ colour-magnitude diagram (CMD) for the Milky Way globular cluster M13. The main features of the CMD are indicated : main-sequence (MS), main-sequence turn-off (MSTO), subgiant branch (SGB), red giant branch (RGB), asymptotic giant branch (AGB), red-side horizontal branch (RHB), RR Lyrae variables (RR), blue horizontal branch (BHB), extreme blue horizontal branch (EBHB), blue stragglers (BS) and white dwarfs (WDs).  Photometric data courtesy of A.R.Gonzales}
\label{fig:CMD}
\end{figure}

A primary tool used to study GCs (and resolved stellar populations in general) is the colour magnitude diagram (CMD). To construct a CMD, images of a GC are obtained in two filters (to construct a colour), and photometry is performed on these images\footnote{The subject of photometry is a chapter in itself. Suffice to say that the standard techniques are reasonably straightforward, although a number of careful steps are required to achieve precise measurements.}.
The CMD for the Milky Way GC M13 is shown in Figure~\ref{fig:CMD}. The different evolutionary phases from the tip of the asymptotic giant branch to the lower main-sequence are shown. A CMD analysis can be used to determine both the cluster age and its metallicity ([Fe/H]).  Some age information is present at a number of  locations in the CMD, although the key age-sensitive  feature is the main-sequence turn-off (MSTO) which indicates the termination  of core H-burning in stars.

The apparent  magnitude of the MSTO depends on distance, so if the distance to the cluster is not well-known, distance-independent measures are generally employed, such as the difference between the position of the MSTO and the horizontal branch \cite{Chaboyer1996,Vandenberg2013}. In contrast, metallicity information comes from the horizontal location of the CMD and also from the shape of the red giant  branch. Increasing metallicity makes the CMD locus shift to the red since more metals in a stellar atmosphere increases  stellar atmospheric opacities, which  results in lower effective temperatures (and redder colours).

To determine ages and metallicities from CMDs isochrones are used. These are theoretical stellar evolutionary tracks which predict stellar temperatures and luminosities (or observationally, magnitudes and colours) for a given age and chemical composition (see Section~\ref{SSPs}).

\subsubsection{Stellar spectroscopy}
\label{StellarSpectroscopy}

The standard technique to study the detailed chemical abundances of stars is high resolution spectroscopy. The definition of "high resolution" is a bit vague, but resolving powers of $R>20,000$ are typical. In order to interpret the spectra of stars, stellar atmosphere models are required. These are theoretical models which predict the strengths of individual atomic and molecular lines for a given effective temperature ($T_{\rm eff}$), gravity ($\log g$) and metallicity. Other factors must also be considered such as "microturbulence", which can also affect the measured line-strengths. In practice, there is a range of stellar atmosphere models which are suitable for different regimes of temperature and chemical abundance. In addition, models may assume "local thermodynamic equilibrium" (LTE) or non-LTE, which can affect the profile of some lines depending upon where they formed in the stellar atmosphere. 

The most basic chemical composition measurement for a star is its metallicity. This generally refers to the ratio of iron to hydrogen ([Fe/H]), since there are many Fe lines in the optical spectra of stars and they are easily measured. However, many other  chemical abundance ratios of individual elements can also be determined, and the stars in GCs show some surprises (this is discussed further in Section~\ref{MilkyWay}).

\subsubsection{Space motions of globular clusters}
\label{SpaceMotions}

Stellar spectroscopy provides precise measurements of radial velocities for the stars in GCs. A radial velocity is the component of the velocity of a GC along our line of sight. However, to know the true space velocities of GCs -- that is their true motions through 3-dimensional space --  requires a knowledge of the cluster proper motion\footnote{The apparent movement on the sky of an object compared to a fixed background.} and distance. The {\it Hubble Space Telescope} ({\it HST}) and the {\it GAIA} satellite have revolutionised the data available for Galactic GCs by  providing proper motions and some parallaxes\footnote{Parallax is the apparent change of position on the sky of an object when viewed from two different positions along a given baseline. If the length of the baseline is known, by measuring the parallax angle (i.e., how much the star appears to move), the distance to the star can be determined.} for the  Milky Way GC system \cite{Vasiliev2019}. The "true space velocities" for 150 GCs are now known, giving important insights into the orbits of the GCs. For example, a recent analysis  for total mass of the Milky Way, based on Milky Way proper motions from the second data release of {\it GAIA} gives $M_{\rm tot}=1.28_{-0.48}^{+0.97}\times 10^{12}$\,M$_\odot$ \cite{Watkins2019}.

\subsection{Spatially unresolved methods : integrated light}
\label{SpatiallyUnresolved}

The  observational techniques used to study spatially unresolved GCs are not too different from resolved stellar studies. However, the analysis of integrated-light information requires different approaches to that of  resolved stellar studies. The integrated colours or spectra obtained for GCs generally captures the light from the entire stellar population, and the interpretation of these observations usually requires simple stellar population modelling.

\subsubsection{Stellar population models}
\label{SSPs}

Simple stellar population (SSP) models require three primary ingredients: a set of isochrones, a stellar library and an assumption about the stellar initial mass function (IMF). The isochrones -- which are invariably theoretical -- predict effective temperatures, surface gravities  and  luminosities for a star of a given age and chemical composition. This isochrone is then populated with stars based on the adopted IMF. The stellar libraries may be empirical or theoretical and both have their advantages and disadvantages. Models made with empirical libraries are generally more successful in reproducing the colours and spectra of GCs and galaxies since they incorporate real stars present in the Milky Way or in GCs. However, this is also a disadvantage. Empirical libraries are limited to those stars observable in the Milky Way, which limits the range of metallicities available (particularly at the metal-rich end) and also imposes a fixed abundance pattern (e.g. [Mg/Fe], [Ca/Fe]) onto the stellar library. For empirical libraries, the parameter space ($T_{\rm eff}$, $\log g$, [Fe/H], [Mg/Fe]...) is generally sampled in a non-uniform manner and special interpolation techniques are implemented (e.g., \cite{Vazdekis2010}).

Theoretical libraries, made from stellar atmosphere models, have no such restrictions. In principle they can be constructed for a wide range of $T_{\rm eff}$, $\log g$, [Fe/H] and with arbitrary abundance patterns. However, they also suffer from several serious drawbacks. For example, no single set of stellar atmosphere models covers the full range of effective temperatures required to model a wide range of ages and metallicities in SSPs. In addition,  the  "linelists" which go into the stellar models are often incomplete and vary between workers. Any missing lines in the models can result in important differences between the modelled and observed stellar population. For more information on SSP modelling and its applications see (\cite{Bruzual2003,Maraston2005,Schiavon2007,Vazdekis2010,Conroy2013}).

\subsubsection{Integrated colours}
\label{Integratedcolours}

The resulting SSPs predict colours and spectra for a simple stellar population for a given age, metallicity and chemical composition. Broad-band, optical colours (e.g., $B$, $V$, $g$, $i$...) can give useful information on ages and metallicities, but they suffer from an age--metallicity degeneracy; more metallic populations look redder, but so do older stellar populations. Disentangling the effects of age and metallicity in the optical can be aided by the use of infrared and near-UV bands for GCs. The colour distributions of GC systems is discussed in Section~\ref{ETGsColours}.

\subsubsection{Integrated spectra}
\label{Integratedspectra}

Integrated spectra are a useful way of breaking the age--metallicity degeneracy. Spectra also provide velocity measurements for GCs, as well as potentially providing information on individual chemical abundance ratios such as the [Mg/Fe]\footnote{Magnesium and iron are produced in short-lived, massive stars which explode as type-II supernovae, whereas iron is produced in longer-lived, lower mass stars which is released in type Ia supernova explosions. Therefore [Mg/Fe] can be used as a chemical clock -- generally the higher the ratio [Mg/Fe], the shorter the timescale of star formation since low-mass stars have not had time to pollute the interstellar medium.} ratio. Individual line-strength indices can be measured and compared to model grids, or an observed spectrum can be compared pixel-by-pixel via "full spectral fitting" techniques.  Metallicities derived in this way, from high quality spectra, provide accuracies of 0.2--0.3\,dex (decades in logarithmic spacing). 

\begin{figure}
\sidecaption
\includegraphics[width=\textwidth]{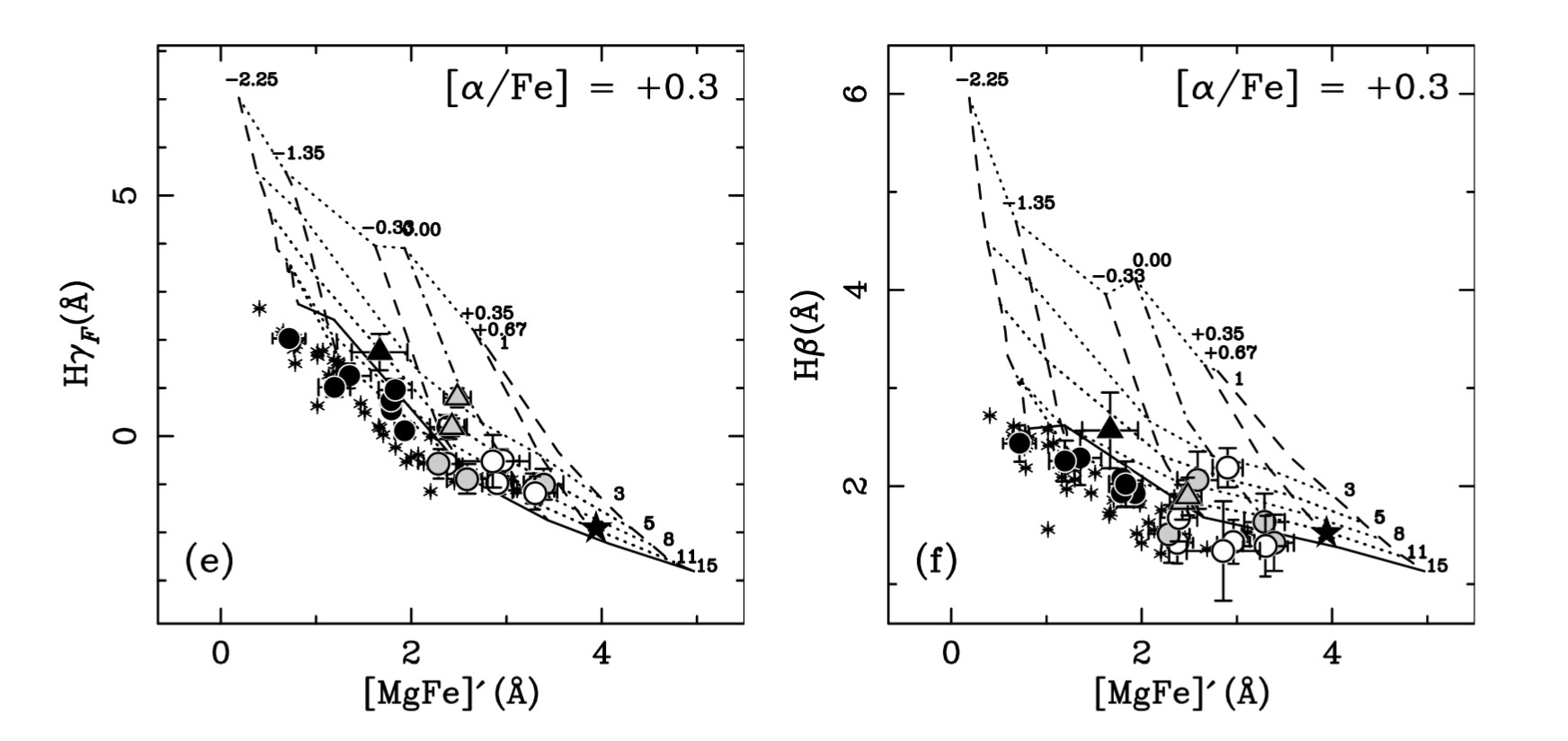}
\caption{Balmer-line versus metal-line from the integrated spectra of globular clusters in the ETG NGC~1407 compared to stellar population models. Near-vertical numbers 1-15 indicate age in Gyr, while the near-horizontal numbers ($-2.25$ to $+0.67$) indicate metallicities. The figure shows that the globular clusters are very old ($\sim14$\,Gyr), and have a wide range of metallicities. Figure taken from \cite{Cenarro2007}.}
\label{fig:Cenarro}       
\end{figure}

Ages can also be determined, in principle, for GCs from integrated spectra. The key age-sensitive lines in the optical integrated spectra of GCs are the Balmer hydrogen series in absorption (H$\alpha$, H$\beta$, H$\gamma$, H$\delta$..) which arise from the electron transitions to the $n=2$ quantum energy level in the hydrogen atom. The Balmer lines are sensitive to temperature and vary in strength with the location (temperature) of the main sequence turn-off. Unfortunately, there are other hot stars in GCs which are not well modeled and also contribute to the Balmer lines. Horizontal branch stars and blue stragglers are particularly problematic. These can lead to ambiguous age determinations, especially for metal-poor clusters. High-quality integrated spectra for GCs in the ETG NGC~1407 \cite{Cenarro2007} are compared to SSP model "grids" in Figure~\ref{fig:Cenarro}.

An outstanding problem with the SSP models is the "zeropoint" problem in GCs (\cite{Gibson1999,Vazdekis2001}). When plotted in a Balmer-line, metal-line diagram, GCs tend to drop off the bottom of the grids below the oldest ages (see right-hand panel of Figure~\ref{fig:Cenarro}). Interestingly, this is generally not seen in spectra for even the oldest galaxies! The origin of this problem is not understood, but it is possible that a combination of the peculiar abundance ratios in GCs (compared to field stars), and also atomic diffusion\footnote{Atomic diffusion is the collective term for processes that change the mixture of atmospheric abundances in stars due to gravity or radiation pressure. For example, heavier elements (e.g., Fe) tend to sink over long time-scales thereby lowering the observed surface abundance of the star.} \cite{Michaud2015} in stars near the turn-off may be responsible \cite{Vazdekis2001}.

\section{The Milky Way globular cluster system}
\label{MilkyWay}

Since the Milky Way GC system is the closest to us, it is also the best studied. A catalogue containing the basic parameters of these GCs is maintained by W.E. Harris \cite{Harris1991} and can be found here\footnote{https://physwww.mcmaster.ca/~harris/mwgc.dat}.
There are presently 158 GCs thought to be associated with the Milky Way. The true number may be closer to $\sim180$ clusters, since a number are probably obscured by the  plane of the Galaxy \cite{Binney2017}. The nearest GC, M4 (NGC~6121), lies just 2.2\,kpc from us. The most distant -- Laevens 1 (also known as Crater) -- is some 145\,kpc away, which puts it nearly three times as far as the Small Magellanic Cloud from which it is thought to originate. A number of GCs are visible to the naked eye. For example, M13 is a fine summer target in dark northern skies and lies just to the west from the centre of the constellation of Hercules.

\subsection{The Origin of the Milky Way globular clusters}
\label{Accreted}

Based on their metallicities, the Milky Way GCs separate into two main populations, a metal-poor population ($\langle$[Fe/H]$\rangle\sim-1.5$) which comprises $\sim$2/3 of the total system, and a metal-rich population ($\langle$[Fe/H]$\rangle\sim-0.5$). A plot of the metallicity distribution function (a histogram of metallicities) looks "bimodal". Spatially, the metal-poor clusters are distributed roughly spherically throughout the Milky Way, reaching out to 145\,kpc into the Galactic halo. The velocities of the metal-poor clusters generally show no ordered motions. However, there are some notable exceptions which might be related to the infall of dwarf galaxies bringing in their own GC populations \cite{Vasiliev2019,Massari2019}. The metal-poor GC system is often referred to as the halo GC population. By contrast, most of the metal-rich clusters lie within 10\,kpc of the Galactic centre. These clusters show a somewhat flattened spatial distribution, and exhibit net rotation of order 50--80\,km/s. The general view is that the metal-rich population is associated with the old "thick" disc or bulge of the Milky Way.

CMD studies of GCs in the Milky Way indicate that the majority of clusters are older than $\sim10$\,Gyr \cite{MarinFranch2009,Vandenberg2013}. That is, the stars in GCs are, in general, at least twice as old as our Sun. In terms of redshift ($z$), this suggests that most Milky Way globulars were formed at $z>2$ whereas our Solar System started formation in the disc of the Milky Way somewhere in the region of $z\sim0.45$. There are, however, a number of clusters which are a somewhat  younger than the majority. Examples of these clusters are Palomar 12 and Terzan 7 which have ages $\sim8$\,Gyr. 

\begin{figure}
\includegraphics[width=\textwidth]{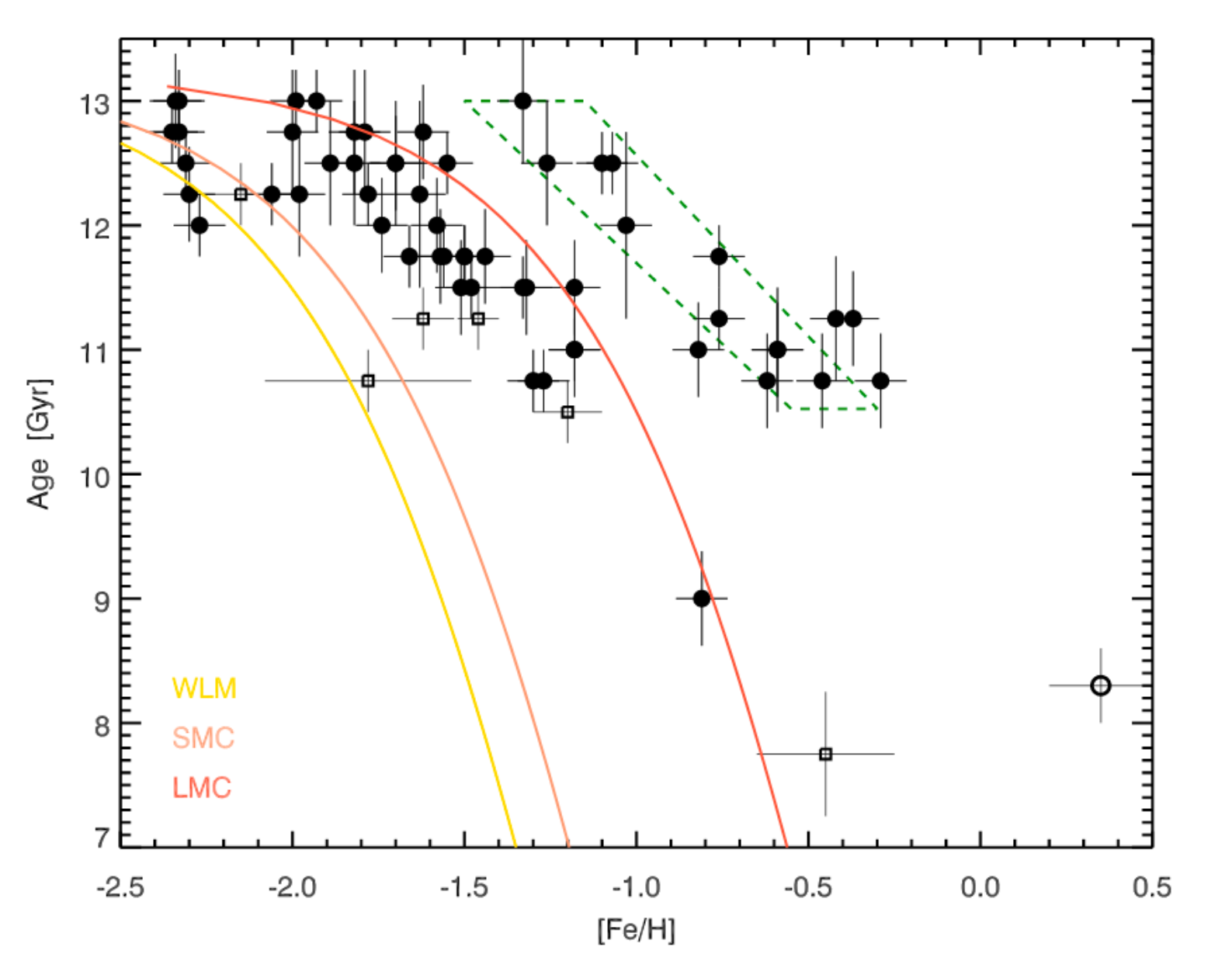}
\caption{Age--metallicity relations (AMRs) for Milky Way GCs (taken from \cite{Leaman2013}). The relations split into several sequences which resemble the AMRs of dwarf galaxies (WLM, SMC, LMC) and also that of the Milky Way bulge (green dashed box). The AMRs suggest that the metal-poor Milky Way clusters may have been accreted from dwarf galaxies.}
\label{fig:Leaman} 
\end{figure}

With increasingly precise ages for Galactic globulars, some very interesting results emerge. When combined with metallicity information, we can plot the age--metallicity relation (AMR) for GCs. This is shown in Figure~\ref{fig:Leaman} and is taken from \cite{Leaman2013}. The figure shows that there are at least two AMRs, and that more metal-rich clusters tend to be younger in any given relation. This is a natural consequence of chemical enrichment during star formation; stars form and the resulting energy, stellar winds and supernovae (stellar "feedback") return metals into the interstellar medium thereby enriching the subsequent generation of stars. However, what is really interesting is that the GCs split into  distinct sequences. I.e., they present a bifurcated AMR. Overplotted are the expected AMRs for several dwarf galaxies (WLM, the Small Magellanic Cloud and the Large Magellanic Cloud). Also shown is the AMR for the Milky Way bulge. The figure shows that the metal-rich clusters ([Fe/H]$>-1.0$) are consistent with having formed along with the bulge of the Milky Way. The bulge is a central, old component of the Galaxy and therefore  these clusters were probably formed along with the main part of the Milky Way. In the context of galaxy formation, this can be referred to as an {\it in-situ}\footnote{Formed "in-place". } GC population.  

In contrast, the metal-poor clusters follow sequences that are consistent with dwarf galaxy AMRs\footnote{This is consistent with what we know about galaxy metallicities. Galaxies follow a stellar mass--metallicity relation in that more massive galaxies are, on average, more metal-rich. This is a consequence of the fact that more massive galaxies have more stars to form metals via nucleosynthesis, and are also better able to hold onto their gas "recycled" from star formation due to their deeper potential wells.}. So why do these GCs look like they formed in dwarf galaxies when they are now in the Milky Way halo? The general conclusion is that these metal-poor "halo" clusters formed in dwarf galaxies and that these dwarf galaxies merged with the Milky Way bringing in their GCs during this process. So, we can regard many of these halo GCs as an accreted or {\it ex-situ} population. The idea that the metal-rich, central Galactic stars and GCs form {\it in-situ}, while (at least an important fraction of) metal-poor halo populations (stars and GCs) are accreted  and represent {\it ex-situ} populations is presently the generally accepted picture for the formation of massive galaxies and their GC systems.

\subsection{Metallicity and abundances}
\label{Metallicity}

\subsubsection{Metallicity Spreads}
\label{Monometallic}

In the Introduction, it was mentioned that GCs are mono-metallic. This is significant because galaxies have spreads in [Fe/H]. Indeed, the presence or absence of measurable spreads in Fe is one proposed way to distinguish galaxies from GCs, similiar to the case for the presence or absence of dark matter (Section~\ref{Whatareglobulars}). Fe is produced in the cores of stars during nucleosynthesis, and can only be released to the interstellar medium via supernova explosions. Therefore, a lack of spread in [Fe/H] implies only one generation of star formation has occurred in most GCs, as opposed to galaxies which may undergo multiple cycles of star formation.

However, it turns out that not all GCs are mono-metallic. A famous example is Omega Centauri (NGC 5139), which turns out to be the most massive GC in the Milky Way with $M\sim4\times10^6$\,M$_\odot$. Spectroscopic studies show that the cluster has a spread in metallicity of up to $\sim1.0$\,dex  and a complex metalicity distribution function (MDF) (e.g., \cite{Norris2012,Johnson2009}). It is now believed that, rather than being a "true" GC, Omega Centauri is actually the nucleus of a dwarf galaxy that accreted into the Galactic halo at some point in the past \cite{Freeman1993,Hughes2000,Bekki2003}. Other GCs with spreads in [Fe/H] include M22 \cite{Marino2011} and M54 \cite{Carretta2010}.

\subsubsection{Light element abundances}
\label{LightElements}

One of the key results to emerge in the past decade has been that GC stars show unusual patterns in some "light" elements (see \cite{Bastian2018} for a review). These elements include He, C, N, Al, Mg, Na and Ca. The picture that is emerging is that there are at least two chemically distinct populations of stars in Milky Way GCs. The first population (P1) shows a pattern of light elements which, for a given [Fe/H], look very similar to field\footnote{Here the term "field" refers to stars not in star clusters.} stars in our Galaxy. The second population (P2) of stars shows a pattern which appears unique to massive star clusters. Specifically, the abundances of the elements H, N, Na and Al are elevated, at a fixed [Fe/H], compared to those seen in P1 and field stars, while C, O and Mg are generally depressed compared to P1 and the field. The origin of these abundance variations is presently unknown, although candidates include winds from rotating massive stars, or material ejected from asymptotic giant branch stars from P1, which somehow may contaminate the (assumed to be) later P2 generation. A hard requirement here is that supernovae from high- or low-mass stars cannot contribute since this would result in a spread in [Fe/H] among the cluster stars which is only seen for a few special cases (Section~\ref{Monometallic}). 

The fact that these unusual light-element patterns are peculiar to GCs might raise the question of why this is interesting in the context of galaxy formation. It turns out the abundance pattern of the P2 generation is sufficiently different from that of Milky Way field stars that it can be used as a chemical "fingerprint" for a number of interesting processes. For example, Galactic stellar surveys suggest that, based on the fraction of "GC-like" P2 stars seen in the Galaxy, somewhere between 10-50\,\% of Milky Way halo stars may actually come from disrupted GCs. \cite{Martell2010,Tang2019}. In addition, it is possible that the abundance variations seen in Milky Way GCs are also responsible for the peculiarities seen in the integrated colours of extragalactic GCs, something that must be understood in order to use the colours of  extragalactic GCs to probe galaxy formation (Section~\ref{ETGsColours}).

\section{Extragalactic globular cluster systems}
\label{Exgal}

For all the detailed information available for the Milky Way GCs, it is only one galaxy. By studying extragalactic systems we can study a wide range of galaxy morphological types and masses in very different environments. This work has lead to important insights into the formation and co-evolution of galaxies and their GC systems.

\subsection{Early-type galaxies}
\label{ETGs}

The GC systems of ETGs have received most attention. This has been in part  due to observational convenience. ETGs have smooth light profiles which makes GCs easily detected. ETGs also tend to have rich GC systems which also helps in their analysis. Beyond observational considerations, ETGs are of great scientific interest. They presently contain more than half of all the stars in the nearby Universe \cite{Fukugita1998}, and they also represent the most massive galaxies known. Analysis of the stellar populations of ETGs indicates that the stars are generally ancient ($\sim10$\,Gyr), metal-rich (equal to or higher than the solar value), and have an IMF dominated by low-mass stars (a "bottom-heavy" IMF) in their centres \cite{Trager2000,Renzini2006,LaBarbera2013}.

\subsubsection{Colours and metallicities}
\label{ETGsColours}

Early ground-based work indicated that the colour distributions of massive ETGs are quite complex and not readily fit by a single gaussian distribution \cite{Zepf1993,Geisler1996}. This lead to a major result of "bimodal" colour distributions in such galaxies, with "blue" and "red" populations. Later works, in particular with {\it HST}, confirmed that the colour distributions of most massive galaxies look bi- or multi-model in optical colours \cite{Kundu2001,Peng2006}. The colour distributions for the most massive galaxies in the centres of clusters can be extremely complex and rather than showing clear bimodality, they appear broad with hints of multiple substructures \cite{Harris2017}. Empirical colour-metallicity relations based on Milky Way GCs, or theoretical relations from SSP models that use standard horizontal-branch recipes, generally predict that such multi-modality in colours maps to multi-modality in metallicity. Detailed spectroscopic studies for a few ETGs with large sample sizes and high-quality spectra generally support the picture of metallicity multi-modality \cite{Strader2007,Beasley2008}\cite{AlvesBrito2011,Usher2012}.

This result of bi-modality was both appealing, since it seems to correspond to the bi-modal metallicity distribution in the Milky Way GC system, but also problematic since producing multi-model metallicity distributions from a modelling point of view proved quite challenging (Section~\ref{Simulations}). Recent investigations have questioned the validity of "one-size fits all" colour-metallicity relations, and the  metallicity distributions inferred from broad-band colours. It has been argued (\cite{Yoon2006}) that the colour-metallicity relations for GCs are non-linear in such a way that intrinsicially unimodal metallicity distributions can be multimodal in colour. This picture is in some disagreement with studies of spectroscopic metallicities (see above) and kinematics (Section~\ref{Kinematics}) and is an area of ongoing research.

Recent studies have also shown that the colour-colour relations for GCs may vary in individual galaxies, and as a function of environment \cite{Powalka2016}. This is a major puzzle, since SSP models predict that for a given age and abundance ratio, any given combination of two colours should trace unique locii as a function of metallicity. Again, this puts into question whether one can use colours as proxies for metallicity for GC systems. The cause of these variations in colour are unknown, but may possibly be related to abundance variations in the GCs, perhaps similar to that seen in the Milky Way GCs (Section~\ref{LightElements}). 

\subsection{Late-type galaxies}
\label{LTGs}

The GCs systems of spiral galaxies have traditionally received less observational attention than the ETGs. This mainly stems from the problems of identifying GCs in imaging with a spatially varying background (spiral arms etc.). In addition, late-type galaxies tend to have less GCs than ETGs for a given mass, and also do not reach the very high masses of the most massive ETGs. A notable exception is that of the  Andromeda galaxy (M31) -- the best-studied GC system of a spiral galaxy with the exception of the Milky Way. M31 has $\sim500$ known GCs, roughly three times the size of the known Milky Way population. Most of these clusters have spectroscopic metallicity estimates, and the metallicity distribution of the clusters looks quite broad, but does not show clear sub-populations in metallicity \cite{Caldwell2011}. The distance of M31 ($d=780$\,kpc) makes secure age determinations from deep CMDs impractical, but the presence of evolved stars (giant branch, horizontal branch) and spectroscopic ages suggest that the majority are old clusters. Wide-area surveys of M31 show numerous structures such as shells and tidal streams which are suggestive of past accretion events (e.g. \cite{McConnachie2009}). A number of the GCs seem to be spatially and kinematically associated with some of these structures offering  evidence for the accretion of GCs onto the halo of this galaxy. 

Beyond the Local Group, several studies have identified what appear to be clusters associated with the discs rather than the halos of their parent spirals \cite{Olsen2004,GonzalezLopezlira2019}. Evidence comes both the spatial distributions of the cluster systems and also signs of ordered rotation whose rotation axis appear consistent with that of the galaxy gas or stellar disc. Since it is unlikely that merging or accretion events can give rise to such disc-like properties, the implication is that GCs can form in the discs of late-type galaxies. Similar conclusions have been reached for the GC system of the Large Magellanic Cloud (see~\ref{Dwarfs}). If the discs of spiral galaxies do contain GCs with ages comparable to Milky Way GCs, then these discs must have been formed at high redshift ($z>2$).

\subsection{Dwarf galaxies}
\label{Dwarfs}

It is important to characterise the properties of the GC systems of dwarf galaxies for a number of reasons, not least being that the progenitors of dwarf galaxies (and their GCs) are thought to build up the halos of more massive galaxies. Most of the more massive dwarf galaxies in the Local Group have GCs and have been studied with both the resolved and unresolved methods mentioned above. There are 14 Local Group dwarfs known with GCs; and the census of these systems continues to grow with increasingly wider-field, high-resolution imaging surveys (see e.g., \cite{Forbes2018a}).

The most massive Local Group dwarf with GCs is the Large Magellanic Cloud (LMC) which has 16 known clusters. CMD and spectroscopic analysis indicates that they are old and metal-poor \cite{Olsen1998,Piatti2019} -- a general property of GCs in dwarfs. Interestingly, the kinematics of these clusters suggest that they might be part of a disc system. This bears similarities to suggestions of "discy" GC systems seen in some spiral galaxies (Section~\ref{LTGs}). At the other end of the mass scale, the Pegasus dwarf has a sole GC located very close to the galaxy centre \cite{Cole2017}. For low-mass dwarfs the properties of their GCs become particularly interesting since they offer the potential of gaining insight into the shape of the dark matter halos of these systems \cite{Cole2012,Orkney2019}.

Beyond the Local Group, GCs have been pivotal in understanding some of the properties of low surface brightness galaxies (the precise definition varies, but these are galaxies typically fainter than the night sky brightness at a dark sight of $\mu(V)\sim22$\,mag\,arcsec$^{-2}$). Recently, there has been a focus on "ultra-diffuse" galaxies (UDGs)\footnote{Not to be confused with "ultra-faint" galaxies which have smaller sizes and significantly lower stellar masses.}, which are a class of low surface brightness galaxy with dwarf-like stellar masses $10^{7-8}$\,M$_\odot$, but large radii ($> 1.5$\,kpc) \cite{vanDokkum2015}. UDGs can have rich GC systems, and velocity measurements of the GCs provided the first dynamical mass measurements for these systems \cite{Beasley2016}(see also Section~\ref{Oddball}). 
 
The dwarf galaxy populations in galaxy clusters are numerically dominated by dwarf elliptical galaxies (dEs). These are spheroidal systems that look superficially like scaled-down versions of ETGs. The origin of these systems is unknown, but is thought to be related to their environment since they are rare in the field. One prominent model is that they are transformed from late-type, gas-rich  dwarfs (dwarf irregulars; dIrrs) via tidal processes \cite{Moore1998,Smith2015}. However, studies of the properties of the GCs in late-type dwarfs and dEs suggest that this model is problematic; dEs tend to have richer GC systems that dIrrs. To solve the problem, new clusters must be formed during the transformation process and these young clusters are generally not seen in dEs \cite{SanchezJanssen2012}.

\subsection{Kinematics}
\label{Kinematics}

Spectra, and therefore radial velocities, of GCs can be obtained for distances out $\sim20$\,Mpc. This includes a number of important clusters and groups such as the Virgo and Fornax clusters, the Leo group and the Centaurus group. Early work on GC kinematics focused on confirming the association of extragalactic GCs with their parent galaxies with radial velocities. It was then quickly recognised that GCs are useful tracers of the  mass distributions of galaxies since they extend beyond the observable galaxy light. Studies of the velocities of GCs in ETGs, spirals and UDGs invariably show the need for dark matter to explain the observations. Typically, the observed random motions of the GCs (their velocity dispersions) is higher than would be expected if only the observed stars and gas contribute to the mass of the galaxy. Along with X-ray studies and gravitational lensing, GCs provide some of the strongest evidence for the presence of dark matter at large radii in ETGs (e.g., \cite{Zepf2000,Cote2001,Strader2011,Zhu2014,Alabi2017}).

GC kinematics also brings useful insights into their connection with their host galaxies. The kinematics of the metal-rich GCs generally looks very similar to that of the central stars in galaxies \cite{Romanowsky2009,Schuberth2010,Strader2011,Norris2012,Pota2013}. On a galaxy-by-galaxy basis, the metal-rich GCs and stars have similar velocity dispersions and rotational properties. This suggests a close relation between the formation processes of the two components. The metal-poor GCs tend to show important differences from the galaxy stars, including differing velocity dispersions, rotation magnitudes (and directions) and also orbital properties. These differences reinforce the picture that the red and blue GCs -- at least in part -- comprise distinct populations, and that their formation pathways are also distinct.

\subsection{Scaling relations between globular cluster systems and galaxies}
\label{Scalingrelations}

GC systems exhibit a number of properties that scale with those of their host galaxies. These relations not only support the idea that galaxy formation and GC formation are intimately connected, but also bring useful insight into these formation  mechanisms. Some of these relations are described below.

\subsubsection{The total numbers of globular clusters in a galaxy}
\label{NGC}

One of the most basic measures of a GC system is the total number of GCs ($N_{\rm GC}$) in the galaxy. Traditionally, the total number of GCs has been estimated by counting GCs up to the peak of the GC luminosity function (GCLF), and then doubling this number \cite{Harris1981}. With this procedure, a symmetrical luminosity function (e.g., a normal distribution) is assumed. It turns out that this is a reasonable approximation, but in detail the GCLF shows more of a tailed distribution known as an evolved Schechter function \cite{Jordan2007} (although not directly relevant to the present discussion, it is also worth noting that the peak GCLF can be used as a "standard candle" to measure distances with an accuracy of about $\sim10$\,\%).

\begin{figure}
\includegraphics[width=\textwidth]{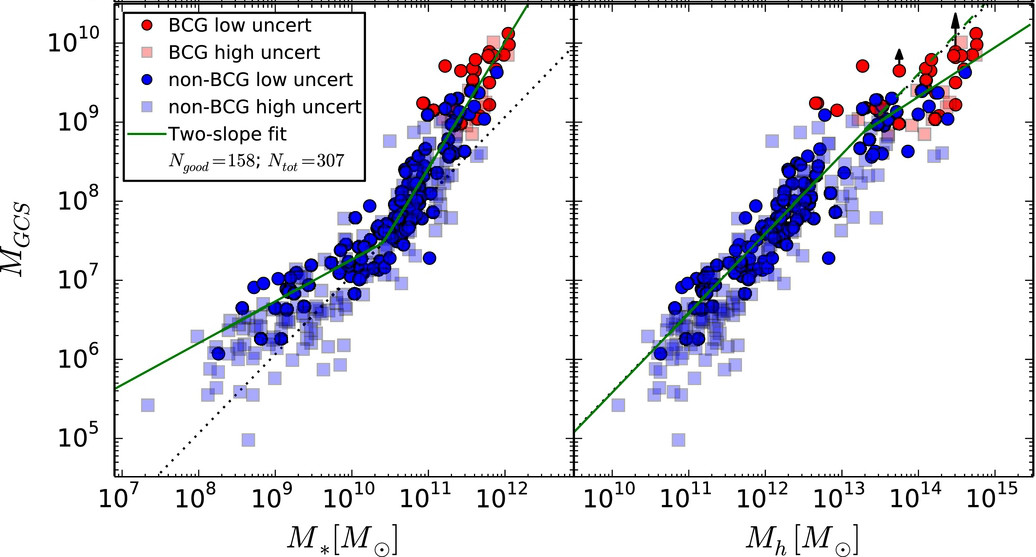}
\caption{Plots of the total mass of globular cluster systems versus parent galaxy stellar mass (left panel) and dark matter halo mass (right-hand panel). The figure (taken from \cite{Hudson2014}) shows that globular cluster systems are connected to the properties of their host galaxies, and show a direct (linear) relation with the dark matter halo of their parent galaxy.}
\label{fig:Hudson}      
\end{figure}

Plotting $N_{\rm GC}$ versus the stellar mass ($M_*$) of the host galaxy shows that $N_{\rm GC}$ correlates positively with $M_*$. By assuming a mass-to-light ratio for the clusters, we can also plot the total GC system mass ($M_{\rm GC}$) versus $M_*$ which is useful since it compares two masses, rather than a number and a mass. This is shown in the left-hand panel of Figure~\ref{fig:Hudson} taken from \cite{Hudson2014}. A positive correlation in $M_{\rm GC}$--$M_*$ may not seem so surprising; the more stars in a galaxy then the more GCs you might have (which are made of stars). However, this result is interesting since it suggests a link between the GC system (total number, or mass) and that of the host galaxy (stellar mass). Further inspection of the figure also reveals that  $M_{\rm GC}$--$M_*$ is not linear. This implies that there is no one-to-one correlation between the two observables. Or, another way to look at it is that the ratio $M_{\rm GC}$/$M_*$ is not constant as a function of mass. It seems that low-mass galaxies (dwarfs) and some very massive galaxies are better at making  GCs than stars than is the case for galaxies of intermediate masses.  

The situation changes when one plots $M_{\rm GC}$ versus the dark matter halo mass ($M_{\rm halo}$) of the host galaxy. Measuring $M_{\rm halo}$ is not straightforward -- we only see the gravitational effects of dark matter not the dark matter itself -- however techniques such as weak lensing can provide meaningful constraints on this quantity (e.g., \cite{Hudson2014}). The right-hand panel of Figure~\ref{fig:Hudson} shows $M_{\rm GC}$ plotted against $M_{\rm halo}$. Unlike the case for $M_*$, this does indicate that there is a linear relation between the two quantities which seems to hold down to at least $M_{\rm halo} \sim 10^{11}$\,M$_\odot$. This implies that there is a constant ratio between the GC system mass and the host galaxy halo mass. At first sight, a linear relation between GC system {\it stellar mass} and host galaxy {\it dark matter mass} might be hard to understand. This result has been used to argue that there might exist a fundamental GC -- dark matter connection. Somehow the GC system of a galaxy "knows" about what sort of dark matter halo it will end up in. The origin of this relation is not fully understood, but a relatively simple explanation may be found in \cite{Elbadry2019}. Essentially, these authors argue that a linear $M_{\rm GC}$--$M_{\rm halo}$ relation emerges as a result of the central limit theorem and galaxy merging; the merging of low- and high-mass haloes and their galaxies leads to "average" halo properties and GC systems which  produce a constant GC to halo mass ratio.

These results above indicate that GCs can be used to trace the properties of the dark matter of their host galaxies. One interesting application is to infer dark matter halo masses for galaxies by measuring $N_{\rm GC}$. This has been done for UDGs (Section~\ref{Dwarfs}). The results of several studies show that typical halo masses for these galaxies are inferred to be $10^{11}$\,M$_\odot$, as derived from $N_{\rm GC}$, which is  consistent with masses from GC dynamics and stellar velocity dispersions.

\subsubsection{Sizes of globular  cluster  systems}
\label{SizeofSystem}

Another GC system property that correlates with the mass of the host galaxy is the size of the GC system \cite{Forbes2017,Hudson2018}. This is typically measured as the "half number radius" (GC $r_{\rm e}$), the radius that contains half of the GC system. GC $r_{\rm e}$ correlates positively with both $M_*$ and $M_{\rm halo}$. 

These correlations are similar to those seen for galaxies; galaxies obey a size--mass relation in that more massive galaxies are, on average, larger. 
Interestingly, galaxies at a fixed stellar mass are more compact at higher redshift. For example, at $z=2$, galaxies with $M\sim10^{11}$\,M$_\odot$ are approximately 3--4 times smaller than nearby galaxies (e.g., \cite{Damjanov2009}). The origin of the "size evolution" of massive galaxies from high redshift to today is not fully understood, but it is believed that mergers play a major role by "puffing-up" compact galaxies, or by adding additional material to their outskirts to make them grow in physical size. It is possible that merging and accretion also give rise to the GC size -- galaxy mass relations reported in the literature.

\subsubsection{The colours of  globular cluster systems}
\label{TheColours}

Studies of ETGs with a range of masses shows that the mean colour (metallicity) of the whole GC system correlates positively with mass of the host galaxy \cite{Strader2005,Peng2006}. More massive ETGs generally have redder GC systems. In addition, the colours of both the red and blue subpopulations scale with galaxy stellar mass. In terms of colour, the relations look a bit different, with the relation for the red clusters being about $\sim5$ times steeper than that for the blue clusters. However, when converted into metallicity, the relations for the both the blue and red clusters appear quite similar \cite{Peng2006}.

The standard interpretation of these observations is that GCs are, on average, able to achieve higher levels of metal enrichment in more massive galaxies. This is similar to the case for the stars in galaxies themselves, which also follow a mass-metallicity relation \cite{Tremonti2004,Kirby2013}.

\subsection{Extreme globular cluster metallicities?}
\label{ExtremeMetallicities}

The maximum and minimum metallicities that a GC achieves offers interesting information on the nature of the interstellar medium (ISM) and star cluster formation early in the history of a galaxy. The most metal-poor globular in the Milky Way is M15 with [Fe/H]$\sim-2.5$\,dex. This is less than 1/300 of the solar value. Based on current data, it seems that a few, if any, extragalactic GCs form with metallicities much lower than this (see \cite{Beasley2019} for a recent data compilation). This "metallicity floor" of GCs is some 5 orders of magnitude higher than the metallicity of the most metal-poor stars known in the Milky Way \cite{Keller2014}. There are a number of interpretations for these observations. Perhaps a low-metallicity ISM is unable to form massive, long-lived clusters due to different fragmentation properties of the gas at these metallicities. Alternatively, it is possible that the sites of GC formation at high-redshift are too low mass at this metallicity to be able to form massive clusters. This latter idea has been developed by \cite{Kruijssen2019}. Understanding the origin of this metallicity floor will bring useful insights into the earliest phases of star and cluster formation.

On the other end of the metallicity scale, the most metal-rich GCs in the Milky Way lie at or near solar metallicities (e.g., NGC 6528 and Pal 10) \cite{Harris1996}. There is no strong evidence that there are significant numbers of GCs in other galaxies with [Fe/H]$>0.0$. This is perhaps surprising for the case of giant elliptical galaxies whose central regions tend to be dominated by super-solar metallicity stars \cite{Trager2000,Gallazzi2005,MartinNavarro2018}. This result might provide a hint that these GCs form relatively early in the star formation process in these galaxies, whose star formation timescales are inferred to be $<1$\,Gyr.

\section{Oddball galaxies and their globular clusters}
\label{Oddball}

In many cases in astronomy, it is the differences from the "normal" population that can give useful insights into underlying astrophysical processes. This is also the case in the study of galaxies and their GC systems. Below a few notable "oddball" systems are highlighted.

\subsection{NGC~4365 - a trimodal globular  cluster system}
\label{N4365}

NGC~4365 is a luminous ETG which lies about 6\,Mpc behind the Virgo cluster of galaxies. Superficially it looks like a normal ETG, although it has a "kinematically decoupled core" (the galaxy centre of the galaxy rotates in the opposite sense to the main galaxy) sometimes taken to be indicative of past merger events \cite{Surma1995}. The galaxy itself, however, appears very old \cite{Davies2001}.

It turns out that the GC system of NGC 4365 is quite unusual. NGC~4365 has been  identified as a galaxy with a "trimodal" colour distribution for its GC system; it has blue, "green" and red GC subpopulations \cite{Puzia2002,Larsen2003,Blom2012,ChiesSantos2012}. The origin of the central green peak has been the subject of some debate in literature, and is still uncertain. One possibility is that the green GCs may result from a gaseous galaxy merger resulting in the formation of new GCs some $\sim4$\,Gyr ago. Alternatively, they may have been stripped (physically removed via gravitational interactions) from the nearby galaxy NGC~4342 \cite{Blom2014}.

\subsection{NGC~1277 - a relic galaxy with only red globular clusters}
\label{N1277}

NGC~1277 is a particularly interesting case of an unusual GC system. NGC 1277 is an extremely old, compact and massive galaxy near the centre of the Perseus galaxy cluster. The galaxy has been identified as a candidate "relic galaxy" which is to say the remnant of the early {\it in-situ} phase in massive galaxy formation without significant subsequent merging or accretion \cite{Trujillo2014}. In this view, NGC~1277 is regarded as a near-pristine "core" of a normal massive ETG. \cite{Beasley2018} explored the GC system with {\it HST} imaging and found that NGC~1277 has few, if any, blue GCs. Since the metal-poor GCs in massive galaxies are generally regarded (at least in part) as a population brought in by the accretion of lower-mass satellites (Section~\ref{Accreted}), an interpretation of the NGC~1277 observations is that the galaxy has undergone very little mass accretion since its formation at high redshift. Somehow NGC~1277 managed to avoid accreting smaller satellite galaxies during its lifetime, possibly due to the fact it is itself a satellite of the more massive  galaxy NGC~1275 which may act to swallow all the material in its vicinity.

\subsection{NGC1052-DF2 -- a galaxy with no dark matter?}
\label{DF2}

"Dragonfly 2" (hereafter DF2; also known as [KKS2000]04) is a low surface brightness dwarf galaxy thought to lie some 20\,Mpc distant in the direction of the massive ETG NGC~1052 \cite{vanDokkum2018}. A study of the kinematics of DF2's GCs suggests that it may have a very low dark matter fraction, compatible with no dark matter at all. This result is intriguing since galaxies generally do have dark matter, and indeed this is one of the definitions of a galaxy (see Section~\ref{Whatareglobulars}). More important than the definition, in modern galaxy formation theory dark matter is generally required in order to produce the galaxies we see around us so DF2 might pose a problem for such models. In addition to an apparent lack of dark matter, the galaxy appears to have an unusual system of GCs in that they are extremely luminous, with a GCLF that is about a magnitude brighter than the usual value ($M_V \sim -7.5$)  (GCs are generally regarded as a standard candle -- see Section~\ref{NGC}).

If DF2 is truly a dark matter deficient galaxy, with a rather peculiar GC system, then it challenges some of our ideas of how galaxies and their GCs can form. However, the galaxy is not without controversy. \cite{Trujillo2019} have argued that DF2 is about twice as close to us as the distance reported by \cite{vanDokkum2018}. This, with additional, different assumptions in their analysis, lead \cite{Trujillo2019} to conclude that DF2 is an ordinary dwarf galaxy with what looks light an ordinary GC system. At the time of writing this Chapter, the jury is out on this one.

\section{Simulating globular cluster  systems}
\label{Simulations}

The improved understanding of star formation and feedback processes, and increasing computing power have allowed for increasingly sophisticated models of GC formation. "Semi-analytic" models and numerical, hydrodynamical simulations have been used to understand the connection between GCs and their host galaxies.

\subsection{Early models}
\label{Earlymodels}

Some of the earliest works to make explicit connections between GC system formation and galaxy formation were largely phenomenological\footnote{I.e., a model that describes relationships between observations, but does not stem directly from physical theory} in nature. \cite{Searle1978} used the mono-metallicity of Milky Way GCs and the lack of a trend in this metallicity with Galactocentric radius (the lack of a metallicity gradient) to argue that the Milky Way halo and its GCs were built-up from "proto-Galactic fragments". This work contrasted quite strongly with an extremely influential paper by \cite{Eggen1962} who considered a smooth, rapid collapse of an early proto-Galactic gas cloud as the precursor to the formation of the Milky Way. It turns out that to some extent these apparently contrasting models have been absorbed in the current picture of galaxy formation. The picture presented by \cite{Eggen1962} looks a lot like {\it in-situ} formation, while \cite{Searle1978} has much in common with {\it ex-situ} accretion.

More specific to the formation of GC systems, \cite{Ashman1992} developed a "major merger" model whereby disk galaxies merge to form elliptical  galaxies, and in the process produce new, metal-rich GCs. These newly formed metal-rich GCs comprise the "red" population, while the "blue" population is brought in as the original GC populations of the merging spirals. This model predicted multi-model colour distributions in elliptical galaxies which were subsequently observed. However, in detail the model had problems. For example it does not explain the sometimes multi-modal colour distributions of spiral galaxies, nor offers an explanation for the origin of the original blue GC populations.
 
Growing recognition that the galactic halo is composed (at least in part) by accreted dwarfs lead to the analytic accretion model of GC formation \cite{Cote1998}. This modelled the build up of cluster systems in massive galaxies via the accretion of lower-mass dwarf satellites. In this model the red population is an {\it in-situ} population of clusters, formed with the galaxy stars, and the blue population is brought in by the accreted satellites. Although the details of GC formation are not described in the model, in spirit this model is consistent with the presently favoured formation pathway for massive ETGs and their GC systems.

\subsection{Semi-analytical models}
\label{SAMs}

The first attempt to place GC formation in a cold dark-matter galaxy formation model was that of \cite{Beasley2002} using so-called "semi-analytic" models (SAMs). SAMs are analytical galaxy formation models that have some of their "free parameters" calibrated based on observations\footnote{In the case of the SAM used by \cite{Beasley2002}, the main calibrations were to match the galaxy luminosity function and Tully-Fisher relations.}. They model the growth and merger history of dark matter haloes as a function of redshift, following the evolution of gas, stars and galaxy formation in these haloes.
\cite{Beasley2002} assumed that GCs formed wherever stars formed, both in gas discs and also during gas-rich major mergers. The model had several observational successes, but in order to produce the colour bimodality a halt to star formation (a "truncation") had to be imposed in the gas discs at high redshift. Possible explanations for this truncation were the re-ionisation of the Universe, or due to the specific pressure conditions in these discs. 
 
\begin{figure}
\sidecaption
\includegraphics[width=\textwidth]{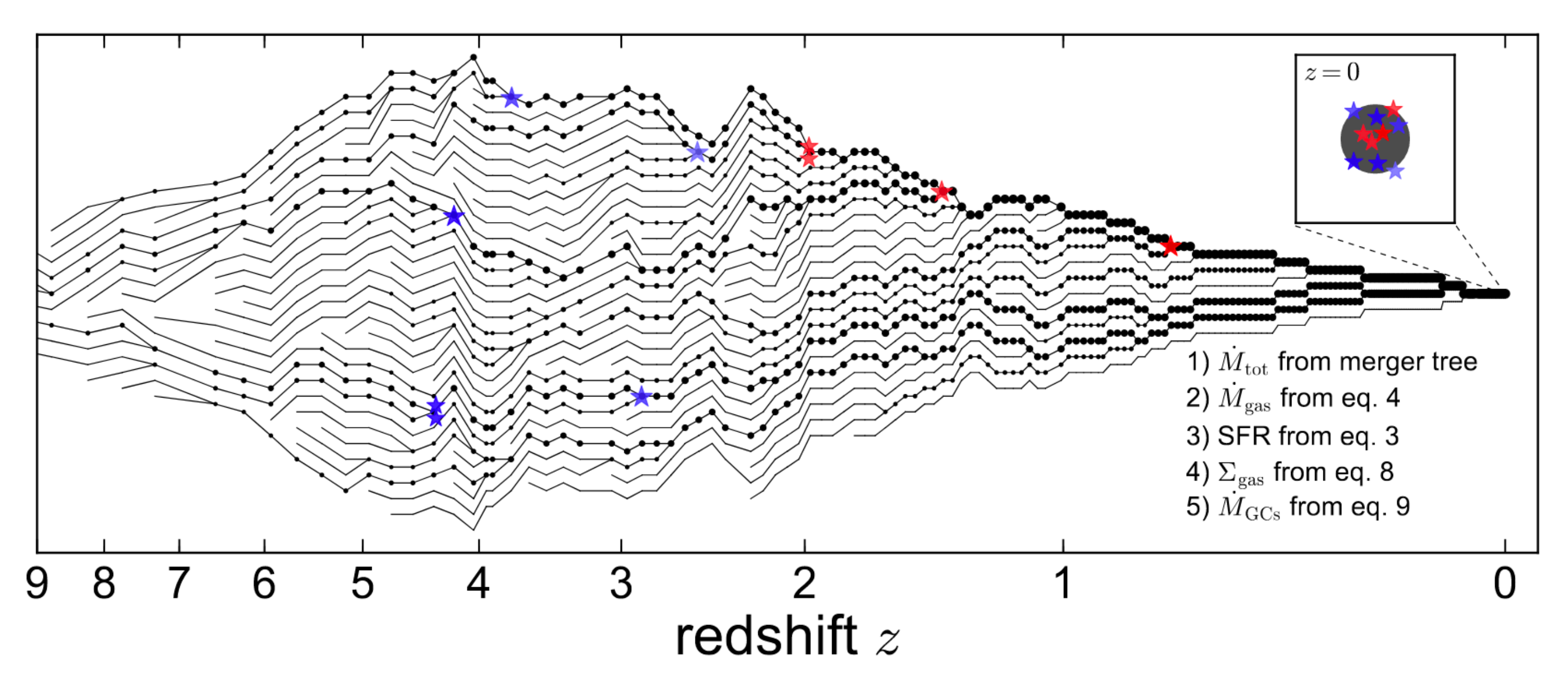}
\caption{Example of a "merger tree" from the semi-analytic models of \cite{Elbadry2019}. The black lines indicate the evolution of individual haloes as a function of redshift merging to become a single galaxy at $z=0$. The blue and red stars indicate globular cluster formation events. The Figure was adopted from \cite{Elbadry2019}.}
\label{fig:ElBadry}       
\end{figure}

More recently, SAMs have been implemented which trace the merger histories of dark matter halos, galaxies and their GCs, but do not always explicitly implement GC formation \cite{Tonini2013,Elbadry2019,Choksi2019}. A "merger tree", showing the growth of galaxy mass via mergers from the SAM of \cite{Elbadry2019} is shown in Figure~\ref{fig:ElBadry}. These works have had success in reproducing some of the GC--galaxy scaling relations such as the $M_{\rm GC}$--$M_{\rm halo}$ relation discussed in Section~\ref{Scalingrelations}. An important advantage of SAMs is that they are computationally fast. Many "virtual" galaxies can be produced in a few minutes on a desktop computer. A potential limitation of SAMs is that they use approximations to calculate various physical processes such as gravitation and merging, gas cooling and star formation. 

\subsection{Numerical Simulations}
\label{NumSims}

An alternative, complementary  approach to SAMs is to use hydrodynamical simulations. These are numerical simulations that follow the evolution of gas and star formation via direct simulation rather than analytic approximations. If the hydrodynamical simulations are "cosmological", then they also follow the evolution of the dark matter component of galaxies based on our understanding of the current cosmological model. 
For example, \cite{Kravtsov2005} simulated a high-redshift, disc-like galaxy and found that gaseous discs at $z\sim3$ are plausible sites for the formation of compact objects that may go on to become present-day GCs.  

Hydrodynamical simulations of galaxy and cluster formation are computationally more expensive than SAMs; a simulation may take weeks, months or even years to run on a supercomputer, depending upon the volume and sophistication of the simulation. However, they have the important advantage of being able to trace the detailed physics of gas cooling, star formation and feedback which is crucial to understanding GC formation. Unfortunately, cosmological, hydrodynamical simulations presently lack the combined spatial (sub-pc scales) and mass resolution ($\sim10^5$~M$_\odot$) to directly model GC formation and evolution down to the present day (i.e., $z=0$).  Because of these computational limitations, "subgrid" recipes (analytic approximations similar in nature to those used in SAMs) are implemented to capture the necessary physics of cluster formation. An example of this is the E-MOSAICS project which use the EAGLE cosmological simulations with specific recipes for GC formation \cite{Pfeffer2018}.
 
\begin{figure}
\sidecaption
\includegraphics[width=\textwidth, angle=270]{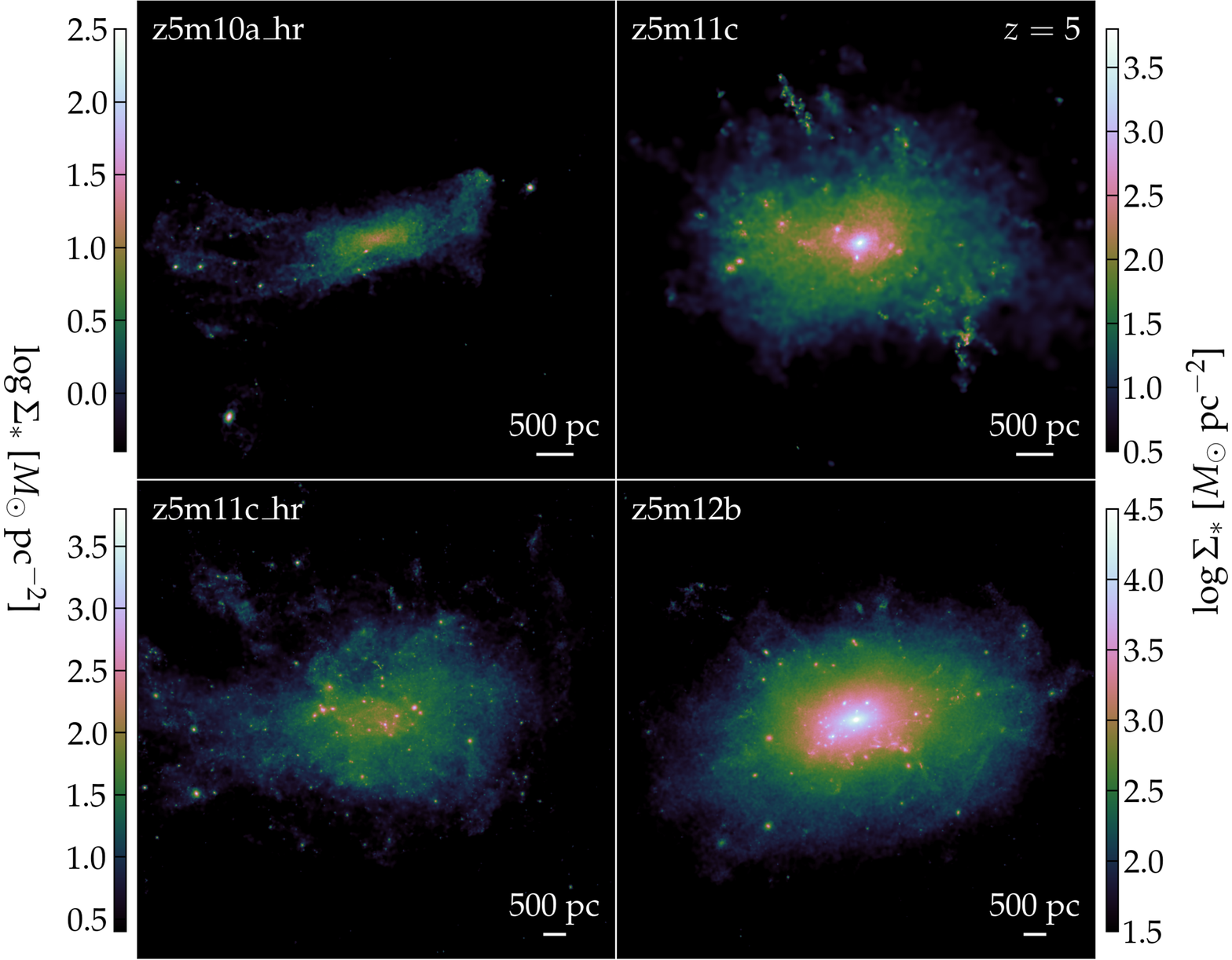}
\caption{Cosmological simulations of globular cluster formation at $z=5$ (taken from \cite{Ma2019}). The colours indicate the density of stars, with lighter colours corresponding to higher densities. Small, point-like objects of high density are star clusters -- possibly young globular clusters.}
\label{fig:Ma}       
\end{figure}

This said, several teams have had recent successes in directly resolving GC formation in simulations that have been limited to high redshifts \cite{Mandelker2018,Kim2018,Ma2019}. Figure~\ref{fig:Ma} shows some "virtual GCs" created in the cosmological simulations from \cite{Ma2019}.

The general picture from these and other works is that GC formation may occur anywhere that is gas-rich and turbulent, such that high pressure regions can form. These regions may be in gas discs, mergers of galaxies, in the very centres of the potential wells of massive proto-galaxies, or perhaps in cold filamentary accretion \cite{Mandelker2018}.
The precise conditions for cluster formation are not well understood, but high gas pressures and densities are probably a key requirement in order to create a compact, bound stellar system \cite{Elmegreen1997}. The above works point to a favoured epoch of GC system formation, which lies somewhere above $z>2$. This is consistent with the ages of the vast majority of the Milky Way GCs (Section~\ref{MilkyWay}).

\section{Globular clusters at high redshift}
\label{Highredshift}
 
The Milky Way GCs are ancient stellar systems (Section~\ref{MilkyWay}), and cosmological simulations place the principal  formation  epoch of GCs at $z > 2$~ (Section~\ref{Simulations}).  Ideally, to test these models one wants to be able to  directly observe the formation of GCs at high redshift, and trace this evolution across different environments and across cosmic time. The greatest challenge to identifying and studying GCs at the highest redshifts is that they are extremely compact. Using special ("PSF-deconvolution") techniques, objects with sizes $\sim100$\,pc can be resolved with {\it HST} at $z\sim6$ (e.g., \cite{Vanzella2019}. A typical,  young GC might have $r_{\rm h} < 20$\,pc and so will be unresolved using standard techniques. Being able to measure the cluster size is crucial, since only a detection in itself will not distinguish a young cluster from (for example) a star forming region in a galaxy, or a compact  galaxy in formation.

An additional problem is one of sensitivity. At $z=6$,  a $10^7$\,M$_\odot$, 10\,Myr old proto-GC will have rest-frame $m_{UV}\sim31$ ($m_V\sim29$). This is beyond the limits of detectability of {\it HST}, but should be within the potential capabilities of the upcoming {\it James Webb Space Telescope} ({\it JWST}) \cite{Renzini2017,Pozzetti2019}.

However, high redshift work studying compact objects has shown that {\it gravitational lensing} already has the potential  to characterise "GC precursors" (GCPs; \cite{Vanzella2019}. In gravitational lensing, the gravitational field of a foreground object (such as a galaxy) acts as a lense which can magnify the apparent size and  increase the brightness of a background source (in this case, a GCP). In strong lensing, the background source may be stretched or show multiple images. Magnification factors may be x10--100, depending upon the precise configuration of the source and lense. In the case of {\it HST}, this allows for effective spatial resolutions of $\sim10$\,pc at $z=6$, well within the expected size range of GCPs.

Several studies have now claimed the possible identification GCPs at high redshift \cite{Vanzella2017,Bouwens2017,Vanzella2019} which can be compared to the properties of GCPs in simulations (Section~\ref{Simulations}). This new area of research opens up the possibility of directly tracing the formation of GC systems when the Universe was a fraction of its present age. Equally as exciting is the prospect of better understanding the process of {\it re-ionization}. At some point early in the lifetime of the Universe (perhaps at redshifts, $z\sim 6 - 20$), neutral hydrogen distributed throughout the Universe was ionized to a plasma by source(s) of energetic (UV) photons. This process ended the so-called "dark ages", the period when no sources of light existed \cite{Barkana2001}. Candidates for the sources of energetic  photons include dwarf galaxies in formation, the first "population III" stars and massive black holes in the centres of galaxies in the form of active galactic nuclei. However, it turns out that GCs, given their short star formation timescales, high redshifts of formation and sheer numbers  (about 2 GCs Mpc$^{-3}$) \cite{BoylanKolchin2018} may be important re-ionization sources. In fact, they may even turn out be the dominant contributors to the UV ionizing background \cite{Katz2014}\cite{BoylanKolchin2018}.

\section{Globular clusters and galaxy formation}
\label{Galformation}
 
The observational and theoretical works on GC systems, a fraction of which has been mentioned in this Chapter, have allowed researchers to build a general picture of the co-evolution of galaxies and their GC systems. Many of the details still need to  be worked out, but a general scenario may be described within the framework of a {\it two-phase} model of galaxy formation \cite{Oser2010}. In this framework, an {\it in-situ} phase builds the centres of galaxies as gas cools and forms stars in dark matter halos at $z>2$. Subsequently, an {\it ex-situ} phase occurs whereby lower mass galaxies are accreted over time to build massive galaxy halos. This accretion phase is still occurring at the present day.

In this context, the formation of GC systems may proceed as follows: The first generations of stars begin to form and then explode, enriching the interstellar medium to metallicities of [Fe/H]$\sim-2.5$, which is the approximate minimum  metallicity seen in GCs. During the {\it in-situ} phase of galaxy formation, in the densest regions of ongoing star formation, the progenitors of today's GCs proceed to form from enriched gas. This formation may proceed in gaseous discs or merging gas-rich systems, but  preferentially occurs near the  centres of what will  become massive galaxies. This produces the red, metal-rich sub-populations of GC systems seen in massive galaxies. In contrast, low metallicity GCs are preferentially formed in low-mass galaxies (dwarf galaxies) which, if accreted onto a more massive galaxy, will go on to form part of the halo (blue, metal-poor) GC population of the galaxy. This represents the {\it ex-situ} phase of GC system formation. In all cases, the formation of individual GCs must occur rapidly (in a few Myr) so as to prevent a second generation of stars creating significant age or metallicity spreads in the majority of GCs, and preferentially occurs at $z>2$.

\section{Summary and outlook}
\label{Summary}

In this Chapter, we have tried to give a taste of some of the research on  GC systems and their connection to galaxy formation. In the case of the Milky Way, detailed ages and abundances of its GCs are increasingly revealing unique information on the formation of our Galaxy and the formation processes of GCs themselves. Further afield, extragalactic GC systems provide information on a range of topics, from the dark matter distributions of galaxies to the mass accretion histories of galaxies  inhabiting a range of extreme environments not represented by the Local Group. Going to high redshift, astronomers are at the point where GC formation can be directly observed, and compared to increasingly sophisticated cosmological simulations. The next generation of telescopes such as {\it JWST} and up-coming 30-m class ground-based facilities (E-ELT, TMT, GMT) will accelerate this rapidly developing, exciting field.

So go and have a look at M13, just to the west from the centre of the constellation of Hercules, and have a think about that.

\begin{acknowledgement}
The author thanks  N\'uria Salvador Rusi\~nol for proof-reading and useful feedback on the text. This work has been supported through the RAVET project by the grant AYA2016-77237-C3-1-P from the Spanish Ministry of Science, Innovation and Universities (MCIU) and through the IAC project TRACES which is partially supported through the state budget and the regional budget of the Consejer\'\i a de Econom\'\i a, Industria, Comercio y Conocimiento of the Canary Islands Autonomous Community.

\end{acknowledgement}

\bibliographystyle{spphys}
\bibliography{references}

\end{document}